\documentclass[12pt,epsf,epsfig,psfig]{article}
\usepackage{graphicx}
\usepackage{epstopdf}
\usepackage{epsfig}
\usepackage{cite}
\def\be{\begin{equation}}
\def\ee{\end{equation}}

\def\lsim{\raise0.3ex\hbox{$<$\kern-0.75em\raise-1.1ex\hbox{$\sim$}}}
\def\gsim{\raise0.3ex\hbox{$>$\kern-0.75em\raise-1.1ex\hbox{$\sim$}}}
\def\NP{{ Nucl.\ Phys.\ }}

\def\PR{{ Phys.\ Rev.\ }}

\def\PRL{{ Phys.\ Rev.\ Lett.\ }}

\begin{document}

%\large

%Draft nutdeco/black \hfill H. Satz, 

%\hfill 25.\ 3.\ 2018

~~\vskip1cm

\centerline{\Large \bf Addendum to}

\bigskip

\centerline{{\Large \bf Strangeness Production and Color 
Deconfinement}\footnote{P.\ Castorina, S.\ Plumari and H.\ Satz,
Int. J. Mod. Phys. E26 (2017) 1750081 
(arXiv:1709.02706)}}

\vskip0.6cm 

\centerline{\bf P.\ Castorina$^{\rm a,b}$, 
S.\ Plumari$^{\rm a,b}$ and H.\ Satz$^{\rm c}$} 

\bigskip

\centerline{a: Dipartimento di Fisica ed Astronomia, 
Universit\'a di Catania, Italy}

\centerline{b: INFN, Sezione di Catania, Catania, Italy}

\centerline{c: Fakult\"at f\"ur Physik, Universit\"at Bielefeld, Germany}

\vskip1cm

\centerline{\large \bf Abstract:}

\bigskip

Recent extensive data from the beam energy scan of the STAR collaboration 
at BNL-RHIC provides the
basis for a detailed update for the universal behavior of the strangeness
suppression factor $\gamma_s$ as function of the initial entropy density,
as proposed in our recent paper \cite{CPS17}.

\vskip1.5cm

This note presents an extension of our recent work on strangeness suppression
in hadronic and nuclear collisions \cite{CPS17}. We had shown there that the
suppression factor $\gamma_s$, needed in hadron resonance gas studies to
account for the production of strange hadrons in $pp$, $pA$ and $AA$ 
collisions at low and intermediate energies, 
becomes a universal function in terms of the initial entropy
density or the initial temperature. Recently, extensive data for hadron
production at various energies and as function of centrality were obtained
by the beam energy scan (BES) of the STAR experiment at RHIC \cite{xu},
and the analysis of these data has provided values for $\gamma_s$
over a considerable range of kinematic variables. We have expressed these
$\gamma_s$ values, given in the experimental analysis as functions of the
collision energy and the centrality, in terms of the initial entropy density
of the corresponding collisions. This is shown to provide an excellent and
more detailed support of our claim that in such a description all $pp$,
$pA$ and $AA$ data fall on one universal curve, which varies from about
0.5 to unity precisely in the color deconfinement transition region.

\medskip

The mentioned BES study of STAR analyses hadron production for Au-Au
collisions at cms energies of 7.7, 11.5, 19.6 27 and 39 GeV, in each
case as function of centrality as measured by the number of participating
nucleons, $N_{\rm part}$, from peripheral to central. The data are subjected
to resonance gas analyses based on grand canonical and canonical (exact
strangeness conservation) formulations, fitting either overall yields or
yield ratios. As a result of these analyses, values of $\gamma_s$ are
provided for $Au-Au$ collisions as function of $\sqrt s$ and centrality.
In our analysis we use the canonical formulation for the
abundance ratios; however, the other forms do not vary greatly. Our input,
as taken from the experimental study, thus starts from the given values of
$\gamma_s(\sqrt s, N_{\rm part})$ for Au-Au collisions.

\medskip

\begin{figure}[h]
\centerline{\psfig{file=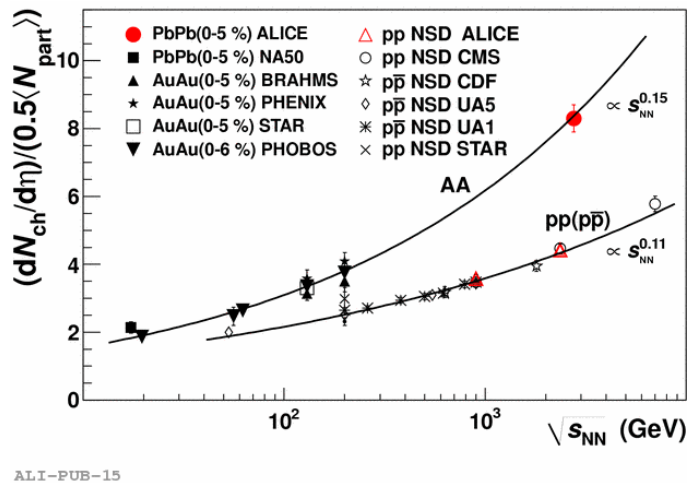,width=9cm}}
\caption{Average charged multiplicities in $pp$ and $AA$ collisions as 
function of cms energy $\sqrt s$ \cite{mult}.}
\label{star1}
\end{figure}

The initial entropy density $s_0$ is given in the one-dimensional hydrodynamic
formulation \cite{Bj} by the form
\be
s_0 ~\!\tau_0 \simeq
{1.5 (N_{\rm part}^x/2) \over \pi R_x^2} \left({dN^x_{\rm ch} \over dy}
\right)_{y=0}^{x},~{\rm with}~ x \sim pp, pA, AA.
\label{star0}
\ee
Here $(dN^x / dy)_{y=0}$ denotes the number of produced charged secondaries,
normalized to half the number of participants $N_{\rm part}^x$, in reaction 
$x$. This form is chosen in multiplicity analyses \cite{mult} in order to 
obtain $(N_{\rm part}^x/2)=1$ for $pp$ and $(N_{\rm part}^x/2)=A$ for central 
$AA$ interactions. The result is shown in Fig.\ \ref{star1}. The fits 
indicated there are provided by \cite{mult}; the details are indicated
also in \cite{CPS17}. 
     
Using this information, we now want to convert the values of 
$\gamma_s(\sqrt s, N_{\rm part})$ into $\gamma_s(s_0 \tau_0)$. This
requires a specification of the transverse area in the different
kinematic situations. As in \cite{CPS17}, we use $R_{pp}=0.8$ fm and
$R_{pPb}=R_p(0.5~N_{\rm part})^{1/3}=1.3$ fm \cite{Abelev,CPS16}; for the
centrality dependent $Au-Au$ collisions, we use $R_{\rm AuAu}=1.25~
(N_{\rm part}/2)^{1/3}$, generalizing the form $R_{AA}=1.25~A^{1/3}$
for central $AA$ collisions to non-central interactions. This is
evidently only a rough estimate; in a more detailed future study, we will
determine the transverse area by means of a Glauber analysis.

\medskip

With the transverse areas thus specified, we can now determine the strangeness
suppression factor $\gamma_s$ as function of $s_0\tau_0$ for the results
of the STAR BES experiments, together with the $pp$ and $pPb$ values
already used in \cite{CPS17}. The result is shown in 
Fig.\ 2, with $s_0 \tau_0$ in units $\rm fm^{-2}$.
In the bottom display, we use a logarithmic scale for $s_0\tau_0$,
in order to show the transition region in more detail. In both cases,
the solid curve shown is
\be
\gamma_s= 1 - 0.997 \exp\{-0.575 s_0\tau_0\};   
\ee
it is meant just to guide the eye and is not a best fit. In \cite{CPS17}, we
had emphasized that the disappearance of strangeness suppression coincides
with the onset of deconfinement. To see this here as well, we recall that
lattice studies \cite{Baza}
give $T_c=154 \pm 9$ MeV for the deconfinement temperature;
the corresponding range for the entropy density is given by 
$s_0 = 2.5 \pm 0.3~{\rm fm}^{-3}$. With the conventional choice of
$\tau_0 \simeq 1$ fm for the thermalization time, Fig.\ 2 corroborates
nicely the coincidence of deconfinement and strangeness equilibration.

\vskip0.7cm

\begin{figure}[h]
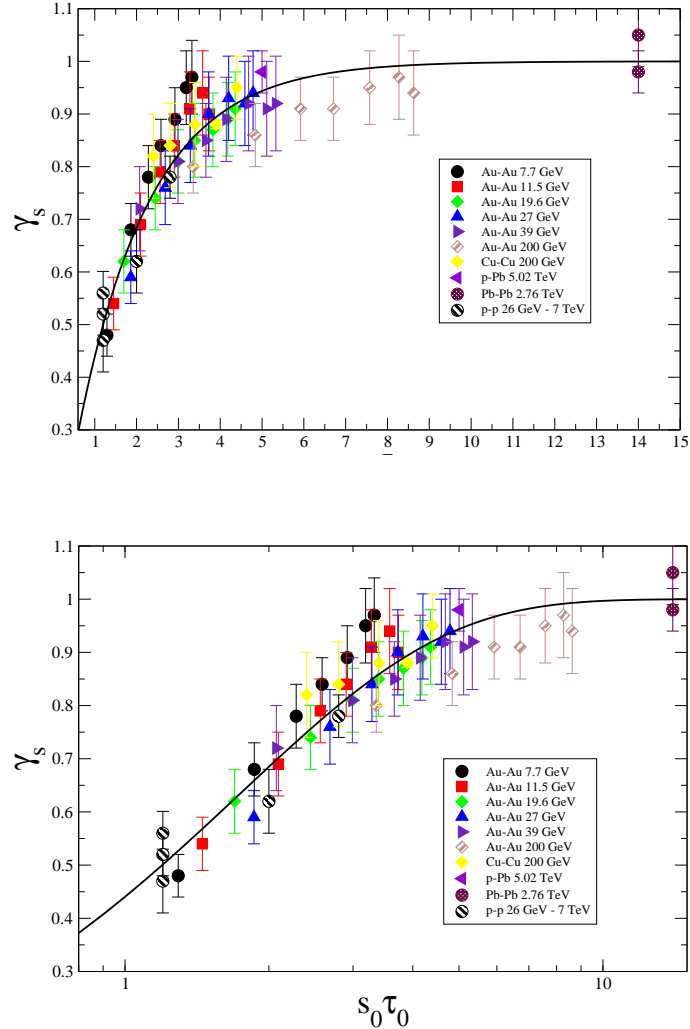

\centerline{\psfig{file=Fstarfinalok.eps,width=9cm}}
\vskip0.65cm
\centerline{\psfig{file=Fstarfinalok1.eps,width=9cm}}
\label{LOK}
\caption{Strangeness suppression factor $\gamma_s$ as function of
$s_0\tau_0$ for $pp$, $pPb$ \cite{ref1}and $Au-Au$ collisions \cite{xu}  
at different collision energies and centralities. For the solid curve,
see text.}
\end{figure}

\vskip0.7cm

From Fig.\ 2 we see moreover that the STAR data at all considered 
collision energies lead for the most central collisions to 
values of 
$\gamma_s$ in the range 0.9 - 1.0. The transition to lower values of
$\gamma_s$ sets in as the collisions become more peripheral, with lower
$s_0\tau_0$. For this reason, various earlier data \cite{early} restricted 
to central collisions had concluded that $\gamma_s \simeq 1$. To obtain a 
complete picture of strangeness suppression in $AA$ collisions and its 
relation to $pp$ interactions, it thus seems necessary to include peripheral 
$AA$ collisions as well.

\vskip2cm

\centerline{\bf Acknowledgements:}

\medskip

We are grateful to Nu Xu (LBL Berkeley/Wuhan) for calling the STAR
data to our attention.
Paolo Castorina gratefully acknowledges support by the
Deutsche Forschungsgemeinschaft (DFG) through the grant CRC-TR 211,
``Strong interaction matter under extreme conditions''.

  \end{document}